\documentstyle [12pt] {article}

\catcode`@=12
\begin{document}
\vspace{0.5in}
\oddsidemargin -.375in
\def\ee{\end{equation}}
\thispagestyle{empty}
\begin{flushright} ISU-NP-96-05\\ISU-HEP-96-05\\SINP-TNP/96-08\\May 1996\
\end{flushright}
\vspace {.5in}
\begin{center}
{\Large\bf $B \rightarrow D(D^*)$ Form Factors 
in a Bethe-Salpeter model
 \\}
\vspace{.5in}
{\bf A. Abd El-Hady${}^a$, 
Alakabha Datta ${}^a$, K.S. Gupta ${}^b$, and J.~P.~Vary ${}^a$\\}
\vspace{.1in}
${}^{a)}$  {\it
 Department of Physics and Astronomy, Iowa State University , Ames, Iowa
50011, USA.}\\
\vspace{.1in}
${}^{b)}$  {\it
Saha Institute of Nuclear Physics, Theory Division, Block -AF, Sector -I,
Bidhannagar,}\\ 
{ \it Calcutta -700064, India } \\ 
\vskip .5in
\end{center}

\vskip .1in \begin{abstract} We calculate the form factors for the
semileptonic decays of the $B$ meson to $D$ and $D^*$ mesons in a
Bethe-Salpeter model. We show that our model is consistent with the
constraints of Heavy Quark Effective Theory (HQET) and we
 extract the matrix elements that represent the $1/m_Q$ corrections
to the form factors in HQET. With available data, we obtain 
$V_{cb} $=$ (31.9 \pm 1.4)\times10^{-3}$.
\end{abstract}
\vskip .25in
\newpage

\section{INTRODUCTION}
In previous papers \cite {BS} we have developed a model for mesons
based on the Bethe-Salpeter equation. Recently \cite {BSV}, we
calculated the Isgur-Wise function $\xi(\omega)$ and extracted the  
Cabibbo-Kobayashi-Maskawa (CKM) 
matrix
element
$V_{cb}$ from data. In this paper we improve upon the work of Ref. \cite
{BSV} by evaluating all the form factors relevant to the semileptonic
decays of the $B$ mesons to the $D(D^*)$ meson. By including additional
$1/m_Q$ effects, we obtain an improved value for $V_{cb}$. 

The discovery of Heavy Quark Symmetry (HQS) in recent years
\cite{H4,H5,H6,H7,H71} has generated considerable interest in the study
of systems containing heavy quark(s). It has been shown that, in the
heavy quark limit, the properties of systems containing a heavy quark
are greatly simplified. HQS results in relations between
non-perturbative quantities, such as form factors, for different
processes involving transitions of a heavy quark to another quark. The
development of Heavy Quark Effective Theory (HQET) \cite {H6} allows one
to systematically calculate corrections to the results of the HQS limit
in inverse powers of the heavy quark mass $m_Q$. Among other
consequences, this has allowed a precision method for determining the
 CKM matrix
elements. An accurate determination of the CKM matrix is crucial for
testing the validity of the Standard Model. 

In spite of impressive results obtained in HQET it has not solved the
problem of calculating the transition form factors in QCD. In
particular, HQS reveals relations between form factors but does not
provide a determination of the form factors. Furthermore, the systematic
expansion of the form factors in $1/m_Q$ in HQET involves additional
non-perturbative matrix elements which are not calculable. We are thus
forced to rely on models for the non-perturbative quantities. However,
the constraints of HQET, which are based on QCD, allows one to construct
models which are consistent with HQET and hence QCD. In this paper we
first
 calculate the form factors in the semileptonic decays of the $B$ meson
to $D$ and $D^*$ mesons in our Bethe-Salpeter model. The parameters of
the model are fixed by fitting the meson spectrum so that the
evaluations of the form factors do not involve any additional free
parameters. The calculated form factors are, therefore, viewed as
predictions of this model. We then obtain $V_{cb}$ from the measured
differential decay rate of $B\rightarrow D^* l \nu$.  We also show that
our model is consistent with the requirements of HQET and that we are
able to extract the unknown matrix elements that appear in the $1/m_Q$
corrections in HQET.

The paper is organized as follows: In Section 2, we discuss the general
formalism for the calculation of the form factors. In Section 3, we
present and discuss the results of our calculations.

\section{THE GENERAL FORMALISM}
The  Lagrangian for the semileptonic decays involving the $b \rightarrow c$
 transition has the standard current-current form after the $W$ boson is
integrated out in the effective theory.  \begin{eqnarray} H_{W} & = &
\frac{G_F}{2 \sqrt{2}}V_{cb} {\bar c}\gamma_{\mu}(1-\gamma_{5}) b
{\bar{\nu}}\gamma^{\mu}(1-\gamma_{5})l \ \end{eqnarray} The leptonic
current in the effective interaction is completely known and the matrix
element of the vector ($V_{\mu}$) and the axial vector ($ A_{\mu}$)
hadronic currents between the meson states are represented in terms of
form factors which are defined in the equations below.  \begin{eqnarray}
\langle D(P_D)| V_\mu |B(P_B) \rangle\ &=&
f_+(P_B+P_D)_\mu+f_-(P_B-P_D)_\mu\nonumber\\ \langle
D^*(P_{D^*},\varepsilon)| V_\mu |B(P_B) \rangle\ &=&
ig\epsilon_{\mu\nu\alpha\beta}\varepsilon^{*\nu}(P_B+ P_{D^*}) ^\alpha
(P_B-P_{D^*})^\beta\nonumber\\ \langle D^*(P_{D^*},\varepsilon)| A_\mu |
B(P_B) \rangle\ &=& f\varepsilon^*_\mu+(\varepsilon^* . P_B)[a_+
(P_B+P_{D^*})_\mu+a_-(P_B-P_{D^*})_\mu] \ \end{eqnarray} $f_{+}, f_{-},
g, f, a_+$ and $a_-$ are Lorentz invariant form factors which are scalar
functions of the momentum transfer $$ q^2 = (P_B -P_D(P_{D^*}))^2$$
where $P_B$, $P_D$ and $P_{D^*}$ are the four-momenta of the $B$, $D$
and $D^*$ mesons respectively.
 The calculation of the the form factors proceeds in two steps. In the
first step, the full current from QCD is matched to the current in the
effective theory (HQET) at the heavy quark mass scale \cite{NR}.
 Renormalization group equations are then used to run down to
a low energy scale $\mu \sim 1 $ GeV where the constraints of HQET
operate and where it is reasonable to calculate matrix elements in a
valence constituent quark model like the one we employ here \cite{ISGW2}.
 
In the HQS limit, the heavy quark inside the
meson acts as a color source and its velocity  remains unchanged 
due to interactions with the
light degrees of freedom. In the leading order the velocity of the heavy
quark is the same as the velocity of the hadron.
 The effect of the external weak current is to instantaneously change
the velocity of the color source from $v$ to a new velocity $v'$. In
Heavy Quark Effective Theory, therefore, it is more appropriate to work
with the velocities of the heavy mesons instead of their momenta and to
consider the form factors as functions of $\omega = v \cdot v'$. The
variable $\omega$ is related to the momentum transfer variable $q^2$
through $$ \omega = \frac{m_B^2+m_{D(D^*)}^2 -q^2}{2m_B m_{D(D^*)}} $$
In terms of a new set of more convenient form factors we can write the
matrix elements of the vector and the axial currents as \begin{eqnarray}
\langle D(v_D)| V_\mu |B(v_B) \rangle\ \over \sqrt{m_Dm_B} &=&
\xi_+(\omega)(v_B+v_D)_\mu+\xi_-(\omega)(v_B-v_D)_\mu\nonumber\\ \langle
D^*(v_D,\varepsilon)| V_\mu |B(v_B) \rangle\ \over \sqrt{m_{D^{*}}m_B}
&=& i \xi_V(\omega)\epsilon_{\mu\nu\alpha\beta}
\varepsilon^{*\nu}v_{D^*}^{\alpha}v_B^{\beta}\nonumber\\ \langle
D^*(v_D,\varepsilon)| A_\mu |B(v_B) \rangle\ \over \sqrt{m_{D^{*}}m_B}
&=& \xi_{A1}(\omega)(v_{D^*} \cdot v_B+1)\varepsilon^*_\mu-
(\varepsilon^*\cdot v_B)[\xi_{A2}(\omega)v_{B\mu}+\xi_{A3}v_{D^{*}\mu}]
\ \end{eqnarray}
 The normalization of the meson states in Eq. (2) and Eq. (3) are
$$\langle M(p')|M(p)\rangle = {2p^0}(2 \pi)^{3} \delta^3(\bf{p-p'})$$
and $$\langle M(v')|M(v)\rangle = \frac{2p^0}{m_{M}}(2 \pi)^{3}
\delta^3(\bf{p-p'})$$

The two sets of form factors defined above are related to each other through
\begin{eqnarray}
\xi_{+} &=& \frac{1}{2}\left(
\sqrt{\frac{m_B}{m_D}}+\sqrt{\frac{m_D}{m_B}}\right)f_+
+ \frac{1}{2}\left(\sqrt{\frac{m_B}{m_D}}-
\sqrt{\frac{m_D}{m_B}}\right)f_- \nonumber\\
\xi_{-} &=& \frac{1}{2}\left(
\sqrt{\frac{m_B}{m_D}}-\sqrt{\frac{m_D}{m_B}}\right)f_+
+ \frac{1}{2}\left(
\sqrt{\frac{m_B}{m_D}}+\sqrt{\frac{m_D}{m_B}}\right)f_- \nonumber\\
\xi_{V} &=& 2{\sqrt{m_D^*m_B}}g \nonumber\\
\xi_{A1} &=& {f \over \sqrt{m_D^*m_B}(v_{D^*}\cdot v_B+1)}\nonumber\\
\xi_{A2} &=& {-m_B^2(a_++a_-) \over \sqrt{m_{D^{*}}m_B}}\nonumber\\
\xi_{A3} &=& {-m_D^*m_B(a_+-a_-) \over \sqrt{m_{D^{*}}m_B}}\
\end{eqnarray}

 In the limit that the mass of the heavy quark $m_Q \rightarrow \infty$
four of the six form factors defined in Eq. (3) can be expressed in
terms of a single form factor, the Isgur-Wise function \cite{H4,H5,H6}
$$ \xi_{+}(\omega) = \xi_{V}(\omega) = \xi_{A_1}(\omega)
=\xi_{A_3}(\omega)
 = \xi(\omega) $$ and the other two simplify to$$
\xi_{A_2}(\omega) = \xi_{-}(\omega) = 0$$
 Furthermore, because of current conservation in the full QCD
Lagrangian,
 the Isgur-Wise function is normalized to unity at zero recoil {\it
i.e.} $\xi (\omega=1)=1$ \cite{Luke}.

 As already noted in the introduction, HQET allows one to systematically 
calculate
$1/m_Q$ corrections to this zeroth order result given above. The 
two sources of the
$1/m_Q$ corrections are from the expansion of the effective Lagrangian
and the quark
fields in the weak currents.
Let us now look into the two sources of $1/m_Q$ corrections.

 In the limit $ m_{Q}\rightarrow\infty $, the
heavy quark field $ Q(x) $ in the full QCD Lagrangian
 is replaced by the effective field $h_{v}(x) $ \cite{H6} 
\begin{equation}
h_{v}(x)  =  e^{im_{Q}v.x}P_{+}Q(x) \\
\end{equation}
where $ P_{+} = {1+\rlap/v\over2} $ is the positive energy projection
operator. The effective Lagrangian with $m_Q \rightarrow \infty$ can be
 written as
\begin{equation}
{\cal L}_{HQET} = \bar h_{v}\,iv\!\cdot\!D\,h_{v}
\end{equation}
where $D^\alpha = \partial^\alpha - ig_{s}t^{a} A^\alpha_a$ is the
gauge covariant derivative.  Corrections to the effective
Lagrangian come from higher dimensional operators suppressed
by inverse powers of $m_Q$.  Including ${1}/{m_Q}$
corrections the effective Lagrangian is \cite{NR}
\begin{eqnarray}
{\cal L} &=& {\cal L} + \delta {\cal L}_1/2{m_{Q}} + ...  \\ \nonumber
\delta {\cal L}_1 & = & \bar{h} \ (iD)^2 \ h + \frac{g_s}{2} \bar{h} \
\sigma_{\alpha \beta} G^{\alpha \beta} \ h
\end{eqnarray}
where $G^{\alpha \beta} = [iD^\alpha, iD^\beta] = ig_{s}t^{a}G_a^{\alpha
\beta}$ is the gluon field strength.
The equation of motion for the heavy quark is
\begin{equation}
v.Dh_v = 0
\end{equation}

Next, one has to  express the currents that mediate the weak
decays of hadrons in terms of the effective field $h_v$.  In 
our case we are interested in
currents of the form $\bar{q} \ \Gamma \ Q$.  At the tree level
the expansion of the current in the HQET takes the form
\begin{equation}
\bar{q} \ \Gamma \ Q \rightarrow \bar{q} \ \Gamma \ h +
\frac{1}{2m_Q} \bar{q} \ \Gamma \ i  \rlap/D \  h + \cdot
\cdot \cdot
\end{equation}
where $\Gamma$ is any arbitrary Dirac structure.
Furthermore, the
hadron mass ($ M_H$) (which appears in the normalization of the states)
 can be expanded in inverse powers of the 
heavy quark mass in
the following manner \cite{NR}.
$$ M_H = m_Q + {\bar \Lambda} + O(1/m_Q) $$
 The mass parameter 
${\bar \Lambda}$ plays a crucial role in the description of $1/m_Q$
corrections to heavy meson and heavy baryon form factors and our
calculation of the form factors will enable us to extract this quantity.
 
In order to calculate the form factors appearing in Eq. (3) we first
match the currents of the full theory, $J^{\mu}_{bc}$, to the current of
the effective theory ${\bf J^{\mu}}_{bc}$.  We can write the matching
condition as \cite{NR} \begin{eqnarray} J^{\mu}_{bc} & = & C_{bc}{\bf
J^{\mu}}_{bc} +\frac{\alpha_s}{\pi}\Delta {\bf J^{\mu}}_{bc} +
\sum_{j}[\frac{B_j}{m_b} + \frac{B'_j}{m_c}]{\bf O}_j + O(1/m_Q^2)\
\end{eqnarray} The second term on the RHS is a new current operator
generated because of operator mixing during the process of matching.
This term has a weak dependence on $\omega$ \cite{NR,ISGW2} and so we
will approximate it with its value at $\omega=1$.  The operators ${\bf
O}_j$ represent the $1/m_Q$ corrections \cite{NR}. Ignoring
$O(\frac{\alpha_s}{m_Q})$ terms, the connection between the QCD
corrected form factors $\xi_{i}$ and the form factors calculated in the
quark model ( {\it i.e.}  in the effective theory) $\xi_i^0$ is
\cite{ISGW2} \begin{eqnarray} \xi_{i} & = & [C_{bc}-1]\xi(\omega) +
\frac{\alpha(\sqrt{m_b m_c})}{\pi}C_{bc} \beta^{\alpha} \xi(\omega)+
\xi_i^{0}(\omega) \ \end{eqnarray}
 where
\begin{eqnarray}
C_{bc}(\omega) & = &
[\frac{\alpha_s(m_b)}{\alpha_s(m_c)}]^{a_I}
[\frac{\alpha_s(m_c)}{\alpha_s(\mu)}]^{a_L(\omega)}
\
\end{eqnarray}
with
$$ a_I = -\frac{6}{33-2N_f}$$
$$a_{L}(\omega) = \frac{8}{33-2N_f'}[\omega r(\omega) -1] $$
$$ r(\omega) = \frac{1}{\sqrt{\omega ^2 -1 }} \ln ( \omega +
\sqrt{\omega^2 -1}) $$
$ \mu$ is  the scale of the effective theory and $N_f=4$, $N_f'=3$. The
expressions for
$$\beta^{\alpha}(\omega)
 \sim
\beta^{\alpha}(\omega =1)$$
are given by \cite{ISGW2}
\begin{eqnarray}
\beta_{+} + \beta_{-} & = & \gamma - \frac{2}{3}\chi \nonumber\\
\beta_{+} - \beta_{-} & = & \gamma + \frac{2}{3}\chi \nonumber\\
\beta_{V} & = & \frac{2}{3} + \gamma  \nonumber\\
\beta_{A_1} & = & -\frac{2}{3} + \gamma  \nonumber\\
\beta_{A_2} & = & -1-\chi + \frac{4}{3}\frac{1}{1-z} +
\frac{2}{3}\frac{1+z}
{(1-z)^2} \gamma  \nonumber\\
\beta_{A_3} & = & -\frac{4}{3}\frac{1}{1-z} -\chi +[{1-\frac{2}{3}
\frac{1+z}{(1-z)^2}}]\gamma  \nonumber\\
z & = & \frac{m_c}{m_b} \nonumber\\
\chi & = & -1 -\frac{\gamma}{1-z}   \nonumber\\
\gamma & = & \frac{2z}{1-z}\ln \frac{1}{z} -2      \
\end{eqnarray}

 The next step is the calculation of the matrix
elements of the currents in the effective theory or, in other words, the
calculation of $\xi_i^0(\omega)$ .
 Such a
        calculation requires the knowledge of the
         meson wave functions. In our formalism the mesons
        are taken as bound states of a quark and an antiquark and the
        meson state is constructed from the constituent quark states.
The wavefunctions 
for the mesons are
calculated  by solving the Bethe-Salpeter equation \cite
{BS} and  include $1/m_Q$ corrections to all orders (in this
particular model). That is, the integral equation carries full
dependence on the finite value of $m_Q$. We
 represent the meson states as \cite{ISGW1}
\begin{eqnarray}
|M({\bf {P_M}},m_J)\rangle\  & = &
\sqrt{2M_H} \int d^{3}{\bf p} \langle L m_{L}S m_{S}|J m_J\rangle\  \langle
s m_s \bar{s} m_{\bar{s}}|S m_S\rangle\ \nonumber\\
 & &\Phi_{L m_L}({\bf p})|\bar q( {m_{\bar
q} \over \ M } {\bf {P}_M} - {\bf {p}},m_{\bar s})
\rangle|q({m_q \over \ M } {\bf {P}_M} + {\bf {p}},m_s)\rangle
\end{eqnarray}
where

\begin{eqnarray}
|q({\bf {p}},m_s)\rangle\ = \sqrt{\frac{(E_q + m_q)}{2m_q}} \pmatrix{
 \chi^{m_s} \cr \ {{\bf{\sigma}}\cdot{\bf p }\over
{(E_q+m_q)}}  \chi^{m_s} \cr }
\end{eqnarray}
 $$M=m_q+m_{\bar q}$$ and $M_H$ is the meson mass.  The meson and the
constituent quark states are normalized as \begin{eqnarray} \langle
M({\bf{P^\prime}_M},m^{\prime}_J)|M({\bf {P}_M},m_J) \rangle\ &=&
2E\delta^3({\bf {P^\prime}_M}-{\bf{P}_M})\delta_{m^{\prime}_J,m_J}
\end{eqnarray} \begin{eqnarray} \langle
q({\bf{p^\prime}},m^{\prime}_s)|q({\bf{p}},m_s) \rangle\ &=&
{E_q\over{m_q}} \delta^3({\bf {p^\prime}}-{\bf
{p}})\delta_{m^{\prime}_s,m_s} \end{eqnarray}
 In constructing the meson states we maintain constituent quark
model approach as we do not include $q\bar{q}$ sea quark states nor the
explicit gluonic degrees of freedom. We also assume the validity
of the weak binding approximation \cite{ISGW1,ISGW2}. 
 In the weak binding limit our meson state forms a representation of the
Lorentz group, as discussed in Ref. \cite{ISGW1}, if the quark momenta
are small compared to their masses. Assuming that the quark fields in
the current create and annihilate the constituent quark states appearing
in the meson state, the calculation then reduces to the calculation of a
free quark matrix element.  In the rest frame of the $B$ meson with a
suitable choice of the four-vector indices in Eq. (3) we can construct
six independent equations which we can solve to extract the six form
factors.

 We now turn to the question of whether our model is consistent with the
requirements of HQET. We will check consistency with HQET up to the
subleading order in $1/m_Q$. We will therefore ignore
$O(1/m_Q^2)$ and higher power corrections even though the wavefunction
from the BSE equation includes power corrections to all order.

Following Neubert and Rieckert \cite
{NRI} we can expand, the form factors $\xi_{i}(\omega) $  including only
$1/m_Q$ corrections as
\begin{eqnarray}
\xi_{i}(\omega) & = & [\alpha_i + \gamma_i(\omega)
+O(\alpha_s^2 , 1/m_Q^2 , \alpha_s/m_Q)]\xi(\omega) \
\end{eqnarray}
  $\alpha_i$ takes the values $0$ or $1$. The
corrections $\gamma_i(\omega)$ represent the $1/m_Q$ corrections. These
corrections are expressed in terms of matrix elements $\rho_i(\omega)$
given below
\begin{eqnarray}
\gamma_{+} & = & (\frac{1}{m_c} + \frac{1}{m_b})\rho_{1}(\omega) \nonumber\\
\gamma_{-} & = & (\frac{1}{m_c} - \frac{1}{m_b})[\rho_{4}(\omega) -
\frac{1}{2}{\bar \Lambda}] \nonumber\\
\gamma_{V} & = & \frac{1}{2} {\bar \Lambda} (\frac{1}{m_c} + \frac{1}{m_b})
+ \frac{\rho_2(\omega)}{m_c} +\frac{[\rho_1(\omega) - 
\rho_4(\omega)]}{m_b} \nonumber\\
\gamma_{A_1} & = & \frac{1}{2} {\bar \Lambda} (\frac{1}{m_c} +
\frac{1}{m_b})\frac{\omega -1}{\omega +1 }
+ \frac{\rho_2(\omega)}{m_c} +\frac{[\rho_1(\omega) 
- \rho_4(\omega) \frac{\omega
-1}{\omega +1}]}{m_b }\nonumber\\
\gamma_{A_{2}} & = & \frac{1}{\omega +1}\frac{[- {\bar \Lambda} + (\omega +
1)\rho_{3}(\omega)
- \rho_{4} (\omega)]}{m_c} \nonumber\\
\gamma_{A_{3}} & = & \frac{{\bar \Lambda}}{2}
 (\frac{1}{m_c}\frac{\omega -1}{\omega +1} + \frac{1}{m_b}) + 
\frac{[ \rho_2(\omega) -
\rho_3(\omega) - \frac{1}{\omega +1} \rho_4 (\omega)]}{m_c}
+\frac{[\rho_1(\omega) - \rho_4(\omega)]}{m_b}\
\end{eqnarray}
 The form factors $\rho_i(\omega)$ can be related to the form factors
$\chi_1(\omega)$, $\chi_2(\omega)$, $\chi_3(\omega)$ and $\xi_3(\omega)$
\cite{FN,TD}
 via
\begin{eqnarray}
\rho_1(\omega) \xi(\omega) & = & \chi_1(\omega) -2(\omega -1)
\chi_2(\omega) +6 \chi_3(\omega)\nonumber\\
\rho_2(\omega) \xi(\omega) & = & \chi_1(\omega) -
2\chi_3(\omega)\nonumber\\
\rho_3(\omega) \xi(\omega) & = &2 \chi_2(\omega) \nonumber\\
\rho_4(\omega) \xi(\omega) & = & \xi_3(\omega) \
\end{eqnarray}
 The normalization conditions at maximum recoil which follows from the
conservation of the vector current in the $m_b=m_c$ limit are \cite{FN}
$$ \xi(\omega=1) =1$$ and $$\rho_1(\omega=1) =\rho_2(\omega =1) =0$$
which is the same as $$\chi_1(\omega=1) =\chi_3(\omega =1) =0$$ Using
the calculated form factors $\xi_i(\omega)$ from the BSE model we can
use Eqs. (18) - Eqs. (20) to extract the HQET parameters ${\bar \Lambda}
$ and $\rho_i(\omega)$or ($\chi_i(\omega)$ and $\xi_3(\omega)$ ). We
note that Eqs. (18) involve 6 equations for 6 form factors. However, the
equations are not linearly independent, so we utilize our model
$\xi(\omega)$ \cite{BSV} and reduce to 5 the number of form factors to
be determined by these equations. We use the least squares method to
solve this system of equations, Eqs.(18), to obtain the HQET
parameters. One should keep in mind that the calculated form factors
contain $1/m_Q^2$ and higher order power corrections
 and so the extracted HQET parameters also contain effects of
$1/m_Q^2$ and higher order corrections.
\section{Results and Discussions}
 In previous papers \cite {BS,BSV} a covariant reduction of the Bethe
-Salpeter equation (BSE) was used to calculate the Isgur-Wise function.
The BSE was solved numerically and the parameters appearing in it(the
quark masses, string tension and the running coupling strength for the
one gluon exchange) were determined by fitting the calculated spectrum
to the observed masses of more than 40 mesons.
The resulting mass spectrum of the analysis was found to agree
 very well with the experimental data. Once the parameters of the model
were fixed, the meson wavefunction could be calculated from the
BSE. This wavefunction was used to calculate the Isgur-Wise function and
determine $V_{cb}$ \cite{BSV}. We now present the results of our present
calculations. A
 number of similar calculations can be found in the literature
\cite{H7,H71}.
For the sake of brevity and whenever appropriate we will only
 compare our results with Ref. \cite{ISGW2}
and the QCD sum rule calculations \cite{H71}.
 
In Fig.1 we show the calculated form factors $\xi_i$ as
function of $\omega$ . We also plot the Isgur-Wise function
$\xi(\omega)$ for comparison. The size of the $1/m_Q$ corrections or
corrections to the HQS limit is
reflected in the deviations of the form factors from
$\xi(\omega)$ or from $0$ 
( in the case $\xi_{-}(\omega)$ and 
$\xi_{A_2}(\omega)$ )
. For  $\xi_{v}(\omega)$, $\xi_{+}(\omega)$, 
$\xi_{A1}(\omega)$ and $\xi_{A3}(\omega)$ the $1/m_Q$ corrections are
positive and can have a maximum effect of $30\%$ without QCD corrections
and about $40\%$ with QCD corrections (See Table 1). Based on the size
of the $1/m_Q$ corrections, we might naively argue that the total 
 $1/m_Q^2$
effects can be expected to be 
about $10-15 \%$. 
For $\xi_{-}(\omega)$ and 
$\xi_{A_2}(\omega)$  
the power corrections are negative and negligible
for $\xi_{A_2}(\omega)$. We could naively
 expect the neglected power corrections
 to follow the same trend as that for the other form factors.  	      
 In this calculation we have not included the
perturbative QCD corrections. 

In Fig.2 we show a plot
of the HQET mass parameter $\bar{\Lambda}$ versus $ \omega$ and we 
see that $\bar{\Lambda}$ is
almost independent of $\omega$ indicating that 
 our model is consistent with
HQET where the mass parameter
 $\bar{\Lambda}$ is independent of $\omega$.  However, as noted before,
our extracted $\bar{\Lambda}$ is modified due to higher power
corrections which are present in the calculated form factors.  The value
of $\bar{\Lambda} \sim 0.55 GeV $ is comparable to the value of this
quantity extracted by other methods, such as QCD sum rules
\cite{PRD,H71}. This contrasts with typical values of $\bar{\Lambda}$
extracted in quark models which are of the order of the constituent mass
of the light quark in the heavy meson (See Ref. \cite{H7} and
Ref. \cite{ISGW2}). We believe that relativistic and spin effects, as
treated in our approach, are responsible for the significant differences
from the traditional quark models.

 In Fig.3 we show the functions that
represent the corrections to the form factors coming from the expansion
of the Lagrangian in $1/m_Q$ 
{\it{viz.}} 
$\chi_1(\omega)$, $\chi_2(\omega)$ 
and $\chi_3(\omega) $ and the correction coming from the
expansion of the
heavy quark field in the weak current
{\it{viz.}} $\xi_3(\omega)$.  
We have plotted the dimensionless quantities
$\chi_1(\omega)/{\bar{\Lambda}}$, $\chi_2(\omega)/\bar{\Lambda}$, 
 $\chi_3(\omega)/\bar{\Lambda} $ 
and 
$\xi_{3}(\omega)/\bar{\Lambda} \xi(\omega)$ in the figure. 

The function 
$\chi_1(\omega)$ represents the
correction coming from the kinetic energy operator while 
$\chi_2(\omega)$, $\chi_3(\omega)$ represent the chromomagnetic
corrections that violate the spin symmetry of the effective theory. 
 At
$\omega=1$, we know from HQET that both $\chi_1$, $\chi_3$ are
  zero. We  expect
  $\chi_1$
and $\chi_3$ to deviate from $0$ at $\omega =1$ because the functions
$\chi_i(\omega)$ includes corrections of
order $1/m_Q^2$ and higher.
In fact the deviation of $\chi_1$ and $\chi_3$ from zero at maximum
recoil is an indication of the size of 
the $1/m_Q^n (n \ge 2)$  corrections. Our results for
$\chi_1(\omega)$, $\chi_3(\omega)$ are close to what one expects in HQET. 
 There are no
constraints on $\chi_2$ at $\omega =1$. We find $\chi_2$ to be small,
positive
and  slowly
decreasing with $\omega$. $\chi_3$ is almost completely flat and remains
close to zero for the entire range of $\omega$ that we have considered.

Our results are consistent with QCD sum rule calculations
\cite{H71} which find the chromomagnetic corrections to be a few
percent. The calculated
$\chi_1$ shows a quadratic behaviour with $\omega$. It peaks around 
$\omega=1.3$ with a maximum value of around $0.22$ which is similar
to QCD sum rule predictions. In HQET the
function
$\xi_{3}(\omega)$ 
is expected to have a $\omega$ dependence similar to the Isgur-Wise 
$\xi(\omega)$. In fact it is customary to write
$$\xi_{3}(\omega)={\bar{\Lambda}}\xi(\omega)\eta(\omega)$$ \cite{NR}
where $\eta(\omega)$ is expected to be a slowly varying function of
$\omega$. Our calculated $\eta$ shows a mild variation with $\omega$
 though the value of $\eta$ 
is smaller than estimated in QCD sum rules. In Ref. \cite{ISGW2}
$\chi_2(\omega)=0$ and $\xi_3(\omega)=0$.
  
The form factors calculated in
our model have to be corrected by taking into account perturbative QCD
corrections which are given in Eqs. (10-13). However, there is an
uncertainty in our choice of the scale $\mu$, the scale of the
effective theory. We
have chosen $\mu
\sim p_{av} \sim 0.6$ GeV 
 where $p_{av}$ is the average value of the internal momentum
inside the mesons \cite{BS}. In Fig.4 we show the form factors
$\xi_i(\omega)$ including perturbative QCD corrections.

 In Table 1 we give the values of the form factors at
$\omega =1$ with and without perturbative QCD corrections. We also show
the numbers calculated in Ref. \cite {ISGW2} for comparison. Even though
we find agreement between our results 
 and those of Ref. \cite{ISGW2} for the sign of the power corrections the
magnitude of the corrections are different. This is probably due to the
wavefunction used in our calculation which includes $1/m_Q$ corrections
 of all order. It is important to note from Table. 1 that the magnitudes 
of the $1/m_Q$ corrections are comparable to the QCD corrections
indicating the need to retain both.

Having obtained
the form factors 
we calculate the decay rate for $B \rightarrow D^* l \nu$ 
 \cite{gs} and fit it to the experimental measurements \cite{CLEO} to
extract $V_{cb}$. The relation for the decay rate in terms of the form
factors is given in Ref. \cite{gs}. The $\chi ^2$ per degree of freedom
of the fit is calculated to be 1.4 without QCD corrections and 0.95 with
QCD corrections.
 
 In Fig.5 we show the decay rate
 calculated using the form factors with and without QCD corrections. In
our most complete approach (including QCD corrections)
we 
extract
a value for $V_{cb}=(31.9 \pm 1.4)\times 10^{-3} $, close to the lower 
limit quoted in the Particle
Data Group \cite{PRD2}. As we have indicated before we might expect about
$10-15\%$ corrections to the form factors from the neglected higher
order power corrections. Since the effects of the $1/m_Q$ corrections
is to bring down the value of $V_{cb}$ from
$(34.7 \pm 2.5)\times 10^{-3} $ calculated in Ref. \cite{BSV}
 to $(31.9 \pm 1.4)\times 10^{-3} $
 we could expect a few percent
corrections to our value of $V_{cb}$ from the neglected $1/m_Q^2$ and
higher power corrections. Since the $1/m_Q^2$ corrections to
the form factors relevant to $B \rightarrow D^* l \nu$ are expected to
be negative \cite{NR} we would expect a higher value for $V_{cb}$ if we
were to include all the $1/m_Q^2$ corrections in our calculations.

	In conclusion, we have presented the form factors in the semileptonic
decays of $B \rightarrow D(D^*) $ in a Bethe-Salpeter model for mesons.
 The parameters of the model are fixed from spectroscopy of the hadrons
and the calculation of the form factors do not involve any new
parameters. We have
 shown that our model is consistent with the requirements of HQET and
have extracted the non-perturbative matrix elements that
characterize the $1/m_Q$ corrections in HQET. We have also obtained
$V_{cb}$ from the available data.

\begin{table}
\caption{The values of the form factors at $\omega =1$}
\begin{center}
\begin{tabular}{|c||c|c|c|}
\hline
$\xi_{i}$ & without QCD corrections & with QCD corrections & ISGW2 
\cite{ISGW2}\\
\hline
$\xi_{+}$ & $1.048$  & $1.086$ & $1.00$ \\
\hline
$\xi_{-}$ &$ -9.37\times 10^{-2}$  &$ -6.94\times10^{-2}$ & $ 
-9.0\times 10^{-2}$ \\
\hline
$\xi_{V}$ & $1.27$  & $1.37$ & $1.17$ \\
\hline
$\xi_{A_1}$ & $1.02$  & $0.99$ & $0.91$ \\
\hline
$\xi_{A_2}$ & $-0.233$  & $-0.120$ & $-0.180$ \\
\hline
$\xi_{A_3}$ & $1.07$  & $1.09$ & $1.01$ \\
\hline
\end{tabular}
\end{center}
\end{table}

{\bf Acknowledgments}
We thank C.Benesh, J.Qiu, J.Schechter, and B.-L.Young for discussions.
This work was supported in part by the US Department of Energy, Grant
No. DE-FG02-87ER40371, Division of High Energy and Nuclear Physics. The
work of A. Datta was supported in part by DOE contract 
number DE-FG02-92ER40730.

\newpage
\section{\bf Figure Captions}
\begin{itemize}
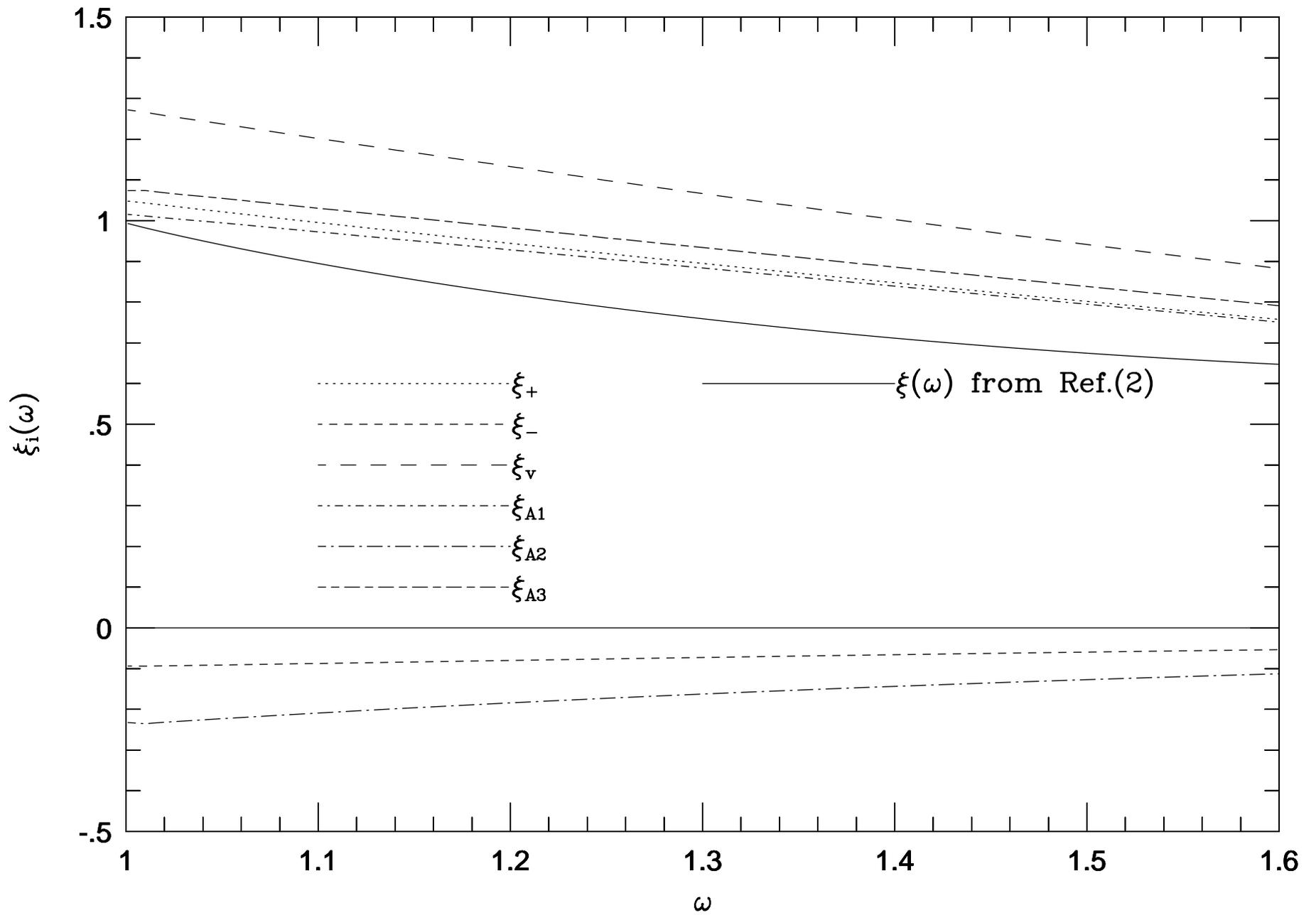
\item[\bf{Fig.1:}] The calculated form factors $\xi_i(\omega)$ without
perturbative QCD corrections and
the calculated Isgur-Wise function $\xi(\omega)$.
 
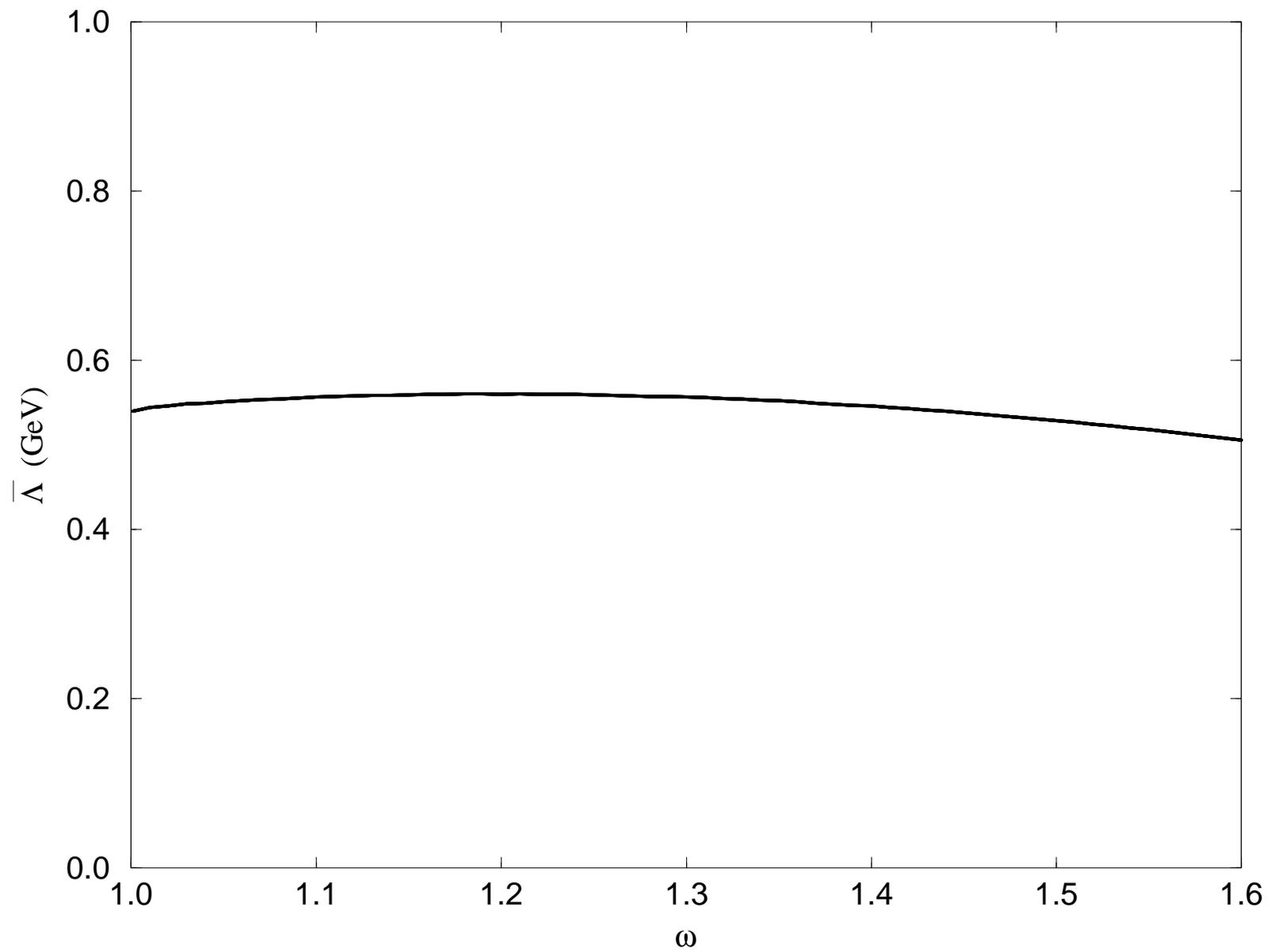
\item[\bf{Fig.2:}] The mass parameter ${\bar \Lambda}$ extracted from
the form factors $\xi_i(\omega)$.

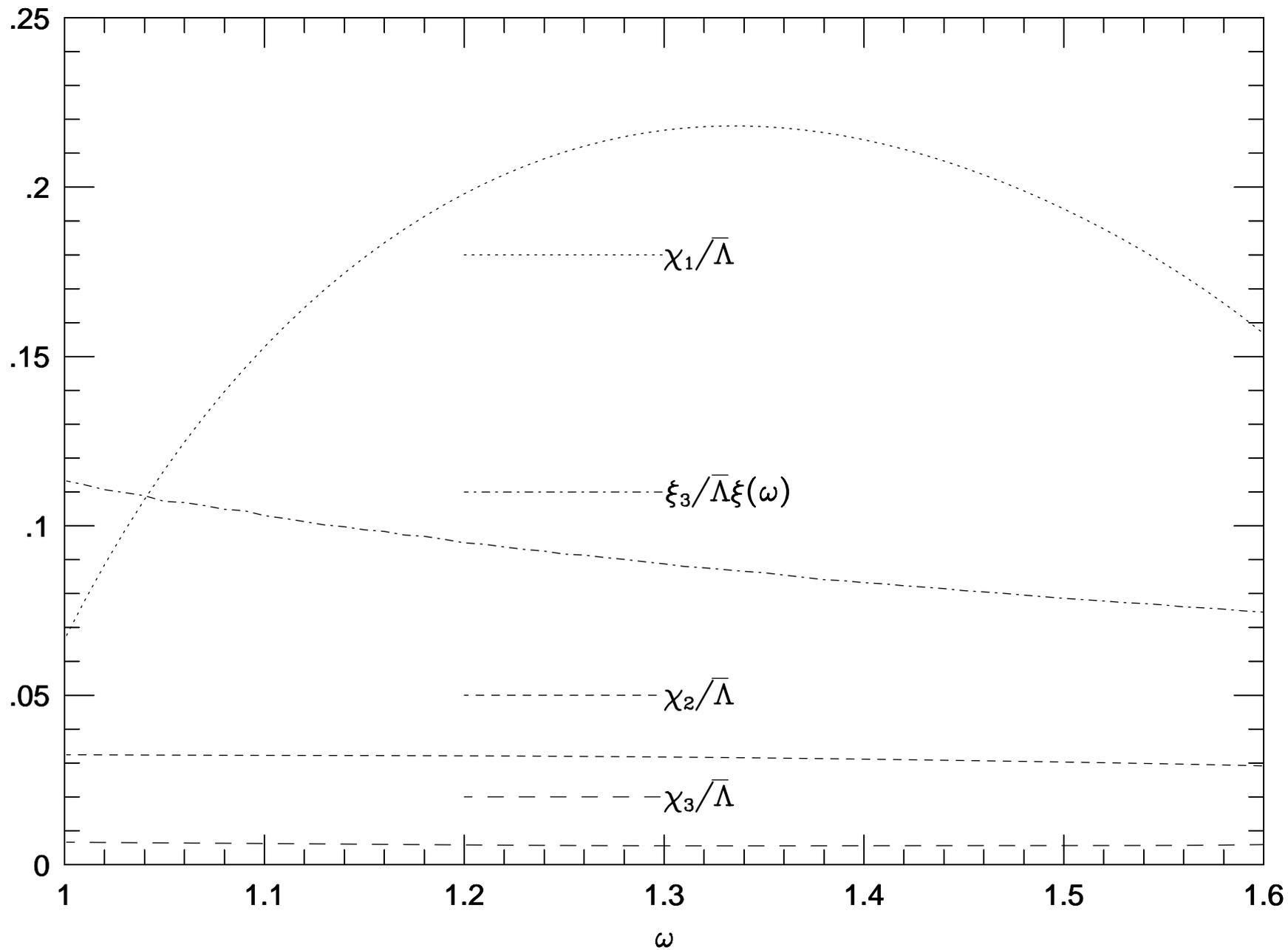
\item[\bf{Fig.3:}] The $1/m_Q $ correction functions
$\chi_1(\omega)/{\bar{\Lambda}}$, $\chi_2(\omega)/\bar{\Lambda}$, 
 $\chi_3(\omega)/\bar{\Lambda} $ 
and 
$\xi_{3}(\omega)/\bar{\Lambda} \xi(\omega)$ 
 extracted from
the form factors $\xi_i(\omega)$. The calculated $\chi_i(\omega)$
and $\xi_3(\omega)$ include
 some $1/m_Q^n(n \ge 2)$ corrections coming
from the meson wavefunctions.

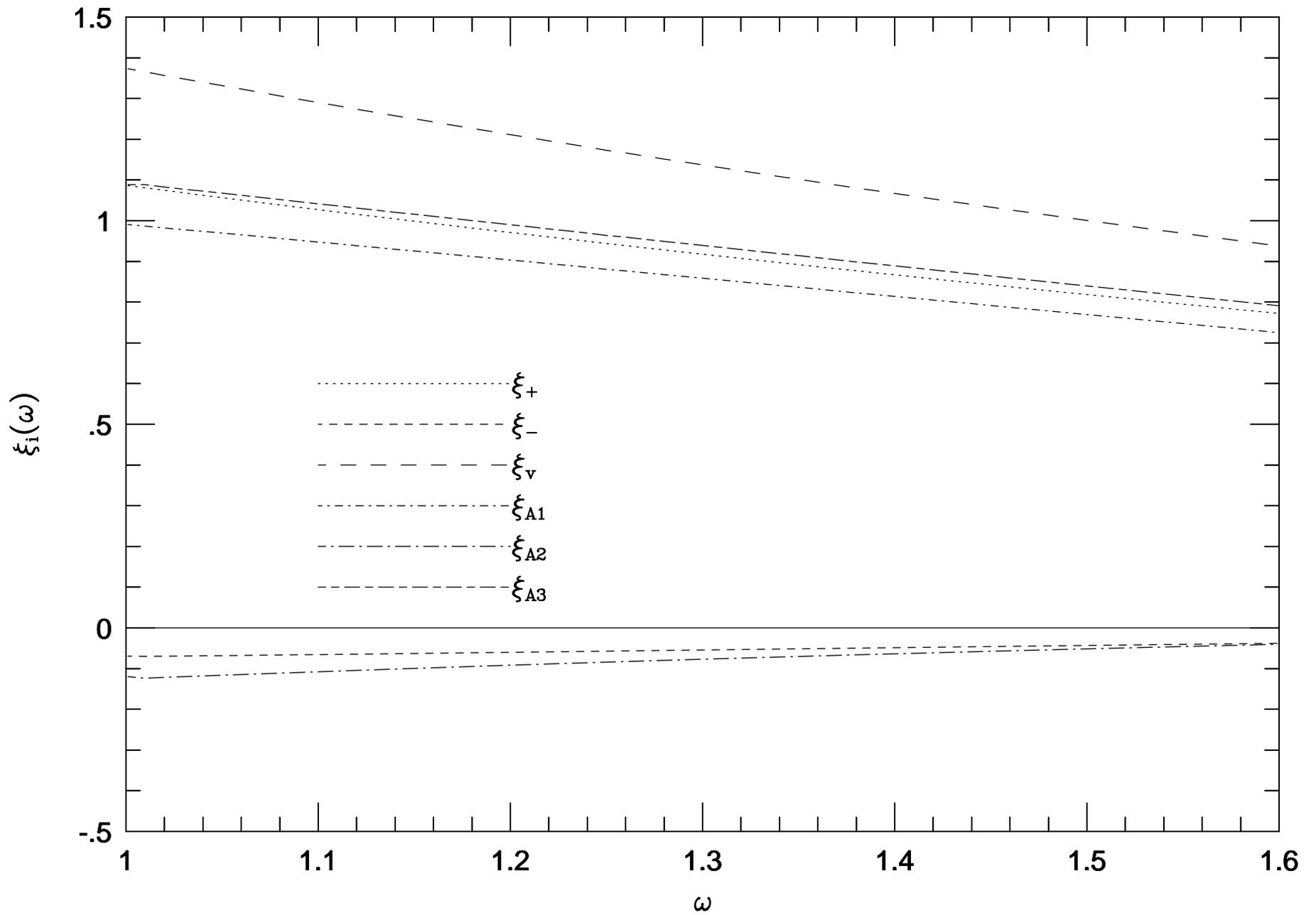
\item[\bf{Fig.4:}] The calculated form factors $\xi_i(\omega)$ with
perturbative QCD corrections with $\mu =0.6 $ GeV.

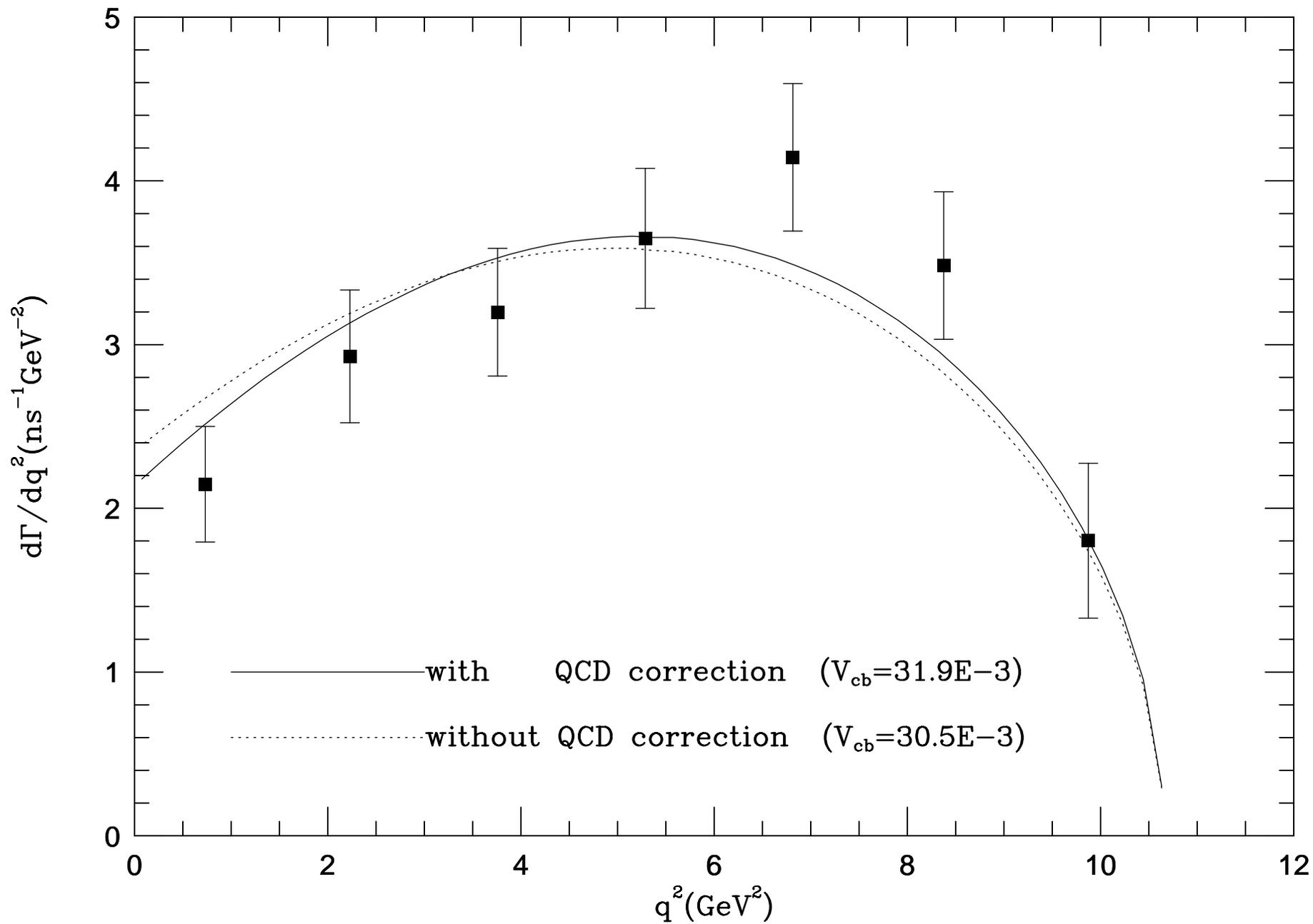
\item[\bf{Fig.5:}] The differential decay rate with and without the QCD
correction, together with the corresponding values of $V_{cb}$.
\end{itemize}

\begin{thebibliography}{References}

\bibitem{BS}
D. Eyre and J.P. Vary, 
Phys. Rev. {\bf D 34}, 3467 (1986);
J.R. Spence and J.P. Vary, 
Phys. Rev. {\bf D35}, 2191 ( 1987);
J.R. Spence and  J.P. Vary, 
Phys. Rev. {\bf C47}, 1282 (1993);
A.J. Sommerer, J.R. Spence and J.P. Vary, 
Phys. Rev. {\bf C49}, 513 (1994);
Alan J. Sommerer, A. Abd El-Hady, John R. Spence, and James P. Vary, 
Phys. Lett. {\bf B348}, 277 (1995). 

\bibitem{BSV}
A. Abd El-Hady, K.S. Gupta, A.J. Sommerer, J. Spence, and  J.P. Vary, 
Phys. Rev. {\bf D51}, 5245 (1995). 

 \bibitem{H4} N. Isgur and M.B. Wise, 
Phys. Lett. {\bf B232}, 113 (1989); 
Phys. Lett. {\bf B237}, 527 (1990);
N. Isgur and M.B. Wise, 
``Heavy Quark Symmetry" in {\it B Decays}, 
ed. S. Stone (World Scientific, Singapore, 1991), p. 158,

{\it ``Heavy Flavors"}, ed. A.J. Buras and M. Lindner 
(World Scientific, Singapore, 1992), p. 234.

\bibitem{H5}
M. B.~Voloshin and M. A. Shifman, Yad. Fiz. {\bf 47}, 801 (1988); 
Sov. J. Nucl. Phys. {\bf 47}, 511 (1988); 
M. A. Shifman in {\em Proceedings of the 1987 International Symposium 
on Lepton and Photon Interactions at High Energies}, Hamburg, 
West Germany, 1987, edited by W. Bartel and R. R{\"u}ckl, 
Nucl. Phys. B (Proc. Suppl.) {\bf 3}, 289 (1988); 
S. Nussinov and W. Wetzel, Phys. Rev. {\bf D36}, 130 (1987);
G.P. Lepage and B.A. Thacker, in {\it Field Theory on the Lattice}, 
edited by A.~Billoire, Nucl. Phys. B (Proc. Suppl.) 4 (1988) 199;
E. Eichten, in {\it Field Theory on the Lattice}, 
edited by A.~Billoire, Nucl. Phys. B (Proc. Suppl.) 4, 170 (1988);
E. Shuryak, Phys. Lett. {\bf B93}, 134 (1980); 
Nucl. Phys. {\bf B198}, 83 (1982).

\bibitem{H6}
H. Georgi, Phys. Lett. {\bf B240}, 447 (1990); 
E. Eichten and B. Hill, Phys. Lett. {\bf B234}, 511 (1990); 
M. B. Voloshin and M.A. Shifman, Sov. J. Nucl. Phys. {\bf 45}, 463 (1987);
H.D. Politzer and M.B. Wise, Phys. Lett. {\bf B206}, 681 (1988); 
Phys. Lett. {\bf B208}, 504 (1988); 
A.F. Falk, H. Georgi, B. Grinstein and M.B. Wise,  
Nucl. Phys. {\bf B343}, 1 (1990);
B. Grinstein,  Nucl. Phys. {\bf B339}, 253 (1990); 
M.B. Wise, ``CP Violation" in {\it Particles and Fields 3:
Proceedings of the Banff Summer Institute (CAP)} 1988, p.~124, 
edited by N.Kamal and F. Khanna, World Scientific
(1989). 

\bibitem{H7}
C.O. Dib and F. Vera, Phys. Rev. {\bf D47}, 3938 (1993);
J.F. Amundson, Phys. Rev. {\bf D49}, 373 (1994);
J.F. Amundson and J.L. Rosner, Phys. Rev. {\bf D47}, 1951 (1993);
B. Holdom and M. Sutherland, Phys. Rev. {\bf D47}, 5067 (1993).

\bibitem{H71}
E. Bagan, P. Ball, V.M. Braun, and H.G. Dosch, 
Phys. Lett. {\bf B278}, 457 (1992); 
M. Neubert, Phys. Rev. {\bf D46}, 3914 (1993); 
M. Neubert, Z. Ligeti, and Y. Nir, Phys. Lett. {\bf B301}, 101 (1993); 
Phys. Rev. {\bf D47}, 5060 (1993).

 
\bibitem{NR}
M. Neubert, Phys. Rep. {\bf 245}, 259 (1994) and references therein;
M. Neubert, CERN-TH/96-55, hep-ph/9604412.

\bibitem{Luke} M.E. Luke Phys. Lett. {\bf B 252}, 447 (1990).

\bibitem{ISGW1}
Nathan Isgur, Daryl Scora, Benjamin Grinstein, and Mark B. Wise, Phys. Rev. 
{\bf D 39}, 799 (1989).

\bibitem{ISGW2}
Daryl Scora and Nathan Isgur, 
Phys. Rev. {\bf D52}, 2783 (1995). 

\bibitem{NRI}
M. Neubert and V. Rieckert, 
Nucl. Phys. {\bf B382}, 97 (1992).

\bibitem{FN}
Adam F. Falk and Matthias Neubert, 
Phys. Rev. {\bf D47}, 2965 (1993).

\bibitem{TD} Debarupa Chakraverty, Triptesh De, Binayak Dutta-Roy and
Anirban Kundu, SINP-TNP-96-04, hep-ph/9603215. 

\bibitem{PRD}
Particle Data Group, 
Phys. Rev. {\bf D50}, 1433-1437 (1994).  
  
\bibitem{gs}
F.J. Gilman and R.L. Singleton,
Phys. Rev. {\bf D41}, 142 (1990).  

\bibitem{CLEO}
B. Barish {\it et al.}, (CLEO)
Phys. Rev. {\bf D51}, 1014 (1995); J.E. Duboscq {\it et al.}, (CLEO) 
CLNS 95/1372.

\bibitem{PRD2}
Particle Data Group, 
Phys. Rev. {\bf D50}, 1315 (1994).  


\end{thebibliography}
\end{document}